\documentclass[12pt,a4paper]{article}
\usepackage{graphicx}
\usepackage[T1]{fontenc}
\usepackage[utf8]{inputenc}
\usepackage{textcomp}
\usepackage[sc,osf]{mathpazo}
\usepackage{a4wide}  
\usepackage{latexsym,amsthm,amsfonts,amsmath,mathrsfs,amssymb}
\usepackage{dsfont}
\usepackage{accents}
\usepackage[nosort]{cite}
\usepackage{booktabs} 
\usepackage[unicode,implicit]{hyperref}
\hypersetup{%
  pdftitle    = {On the generalized Komar charge of supersymmetric solutions}
  pdfkeywords = {gravity,  supergravity, unbroken supersymmetry, conserved charges, Komar charge, Komar integrals, BPS bounds},
  pdfauthor   = {Tom\'as Ort\'{\i}n},
  plainpages  = true,
  colorlinks  = true,
  citecolor   = blue,
  urlcolor    = red,
  linkcolor   = black
}
\newcommand{\hepth}[1]{{\tt
\href{http://www.arXiv.org/abs/hep-th/#1}{hep-th/#1}}}

\newcommand{\arxiv}[1]{{\tt arXiv:\href{http://www.arXiv.org/abs/#1}{#1}}}
\newcommand{\doi}[1]{{\tt DOI:\href{http://dx.doi.org/#1}{#1}}}

\allowdisplaybreaks

\makeatletter
\@addtoreset{equation}{section}
\makeatother

\begin{document}

\begin{flushright}
\small
IFT-UAM/CSIC-26-061\\
May 7\textsuperscript{th}, 2026\\
\normalsize
\end{flushright}

\vspace{1cm}

\begin{center}

  {\Large {\bf On the generalized Komar charge\\[.5cm]
      of supersymmetric solutions}}
 
\vspace{2cm}

\renewcommand{\thefootnote}{\alph{footnote}}

{\sl Tom\'{a}s Ort\'{\i}n}\footnote{Email: {\tt Tomas.Ortin[at]csic.es}}

\setcounter{footnote}{0}
\renewcommand{\thefootnote}{\arabic{footnote}}
\vspace{1cm}

{\it\small Instituto de F\'{\i}sica Te\'orica UAM/CSIC\\
C/ Nicol\'as Cabrera, 13--15,  C.U.~Cantoblanco, E-28049 Madrid, Spain}

\vspace{3cm}

{\bf Abstract}
\end{center}
\begin{quotation} {\small We consider the supersymmetric solutions of
    several supergravity theories (ungauged $\mathcal{N}=2,d=4$ and
    $\mathcal{N}=1,d=5$ supergravities coupled to vector
    supermultiplets and pure $\mathcal{N}=1,d=10$ supergravity) and
    show that their generalized Komar charges vanish identically for
    the supersymmetric Killing vector which is constructed as a
    bilinear of the Killing spinors of those solutions. This property
    can be used to prove the supersymmetry (``BPS'') bounds satisfied
    by some of those solutions in a rigorous, coordinate-independent
    way.}
\end{quotation}

\newpage
\pagestyle{plain}

\tableofcontents


\section{Introduction}

Both from the conceptual and the merely utilitarian points of view,
the conserved charges of physical systems are the most important tools
we can use to characterize their states and their evolution, both at
the classical and quantum levels. It is for this reason that there is
a permanent interest in identifying and defining new conserved charges
and in finding the laws that dictate or constrain their evolution.

The concept of conserved charge has also evolved with time to
accommodate gravitational or non-Abelian Yang--Mills charges for which
there is no local conservation law, that is: no density $(d-1)$-form
current $\mathbf{J}$ satisfying the continuity
equation\footnote{$\mathbf{J}$ is nothing but the Hodge dual of the
  1-form current $j_{\mu}dx^{\mu}$ and Eq.~(\ref{eq:dJ=0}) is
  equivalent to $\nabla_{\mu}j^{\mu}=0$.}

\begin{equation}
  \label{eq:dJ=0}
  d\mathbf{J}=0\,.
\end{equation}

Instead, gravitational and non-Abelian Yang--Mills charges are
expressed as surface integrals at infinity of $(d-2)$-form ``charges''
$\mathbf{Q}$ which, asymptotically, satisfy a similar equation
\cite{Abbott:1981ff,Abbott:1982jh}

\begin{equation}
  \label{eq:dQ=0}
  d\mathbf{Q}=0\,.
\end{equation}

The surface integrals of these $(d-2)$-form charges give the total
charges of the Universe as the only conserved quantities.

Somewhat surprisingly, in some particular cases, it is possible to
define $(d-2)$-form charges that satisfy the above equation everywhere
and whose integral gives the same result. For lack of a better name,
we are going to call them \textit{generalized Komar charges}. The
property Eq.~(\ref{eq:dQ=0}) extended over the whole spacetime tells
us that same result is obtained when we integrate over any other
surface that we can obtain by a smooth deformation of the surface at
infinity without crossing any sources. While this is just a
generalization of Gauss' law of electromagnetism, it is usually
referred to as ``conservation'' and Eq.~(\ref{eq:dQ=0}) is usually
referred to as a ``conservation law,'' and the same expressions are
now used for forms of any other ranks which are closed on- or
off-shell.

In theories of gravity, possibly coupled to matter fields, one can
construct generalized Komar charges for solutions which are invariant
under the general coordinate transformation generated by a vector
field $k=k^{\mu}\partial_{\mu}$. Since the metric is one of the fields
left invariant, $k$ must necessarily be a Killing vector field.  We
will denote them by $\mathbf{K}[k]$ and, by definition, they satisfy
Eq.~(\ref{eq:dQ=0}) on-shell, that is

\begin{equation}
  \label{eq:dK=0}
  d\mathbf{K}[k]\doteq 0\,.
\end{equation}

The first and simplest example of these charges is the standard Komar
charge of pure General Relativity \cite{Komar:1958wp}

\begin{equation}
  \mathbf{K}[k]
  =
  -\frac{1}{16\pi G_{N}^{(d)}} \star d\hat{k}\,,
\end{equation}

\noindent
where $\hat{k}=k_{\mu}dx^{\mu}$. For stationary, asymptotically-flat
solutions and for the timelike Killing vector $k$ normalized $k^{2}=1$
at infinity, its integral over the $(d-2)$-sphere at spatial infinity
$S^{d-2}_{\infty}$ gives the ADM mass $M$, up to a
dimension-dependent constant

\begin{equation}
  \label{eq:Komarmass}
  \int_{S^{d-2}_{\infty}}\mathbf{K}[k]  = \frac{(d-3)}{(d-2)}M\,,
\end{equation}

\noindent
but the same result is obtained by integrating over any other surface
as long as we do not cross any singularity.

When there is matter coupled to gravity, the Einstein equations change
and the standard Komar charge is no longer
conserved.\footnote{Nevertheless, the \textit{Komar integral}
  Eq.~(\ref{eq:Komarmass}) is still used to calculate the mass at
  infinity.} Often, however, it is possible to find a generalization
of the Standard Komar charge that includes more terms with the
property Eq.~(\ref{eq:dK=0}). For the Einstein-Maxwell-cosmological
theories, these generalizations were found in
Refs.~\cite{Bardeen:1973gs,Carter:1973rla,Magnon:1985sc,Bazanski:1990qd}
and, for more general theories an algorithm based on the Noether
theorems and on the observations of Ref.~\cite{Liberati:2015xcp} has
been proposed and elaborated in
Refs.~\cite{Ortin:2021ade,Mitsios:2021zrn,Meessen:2022hcg,Meessen:2022hcg,Ortin:2022uxa,Ortin:2024mmg,Cerdeira:2025elp}.

With an adequate asymptotic normalization of the elements that
contribute to the generalized Komar charge of most theories (in
particular making the electrostatic potentials vanish at infinity) one
can recover Eq.~(\ref{eq:Komarmass}) but, in general, the Komar
integrals at infinity contain contributions from other conserved
charges (electric, magnetic...) and the structure of the final result
resembles the combinations of mass and charges that are bounded by
below by supersymmetry. These \textit{supersymmetry
  bounds},\footnote{The name BPS bound is also used, although not all
  BPS bounds are related to supersymmetry.}, saturated by
supersymmetric solutions with the right asymptotic behavior, have
played a very important role in many developments in supergravity and
superstring theory, specially in black-hole physics\footnote{For a
  comprehensive review on supersymmetric solutions and supersymmetry
  bounds see, for instance, Ref.~\cite{Ortin:2015hya}. A recent
  review on supersymmetric and non-supersymmetric black-hole solutions
  in supergravity and superstring theories is~\cite{Ortin:2024slu}.}

Indeed, it is not difficult to convince oneself that supersymmetry
bounds must arise as results of the Komar integrals.  Generalized
Komar charges are used to derive Smarr relations between a combination
of conserved charges defined at infinity and a combination of
thermodynamical quantities defined on the horizon. The identity
between these two combinations follows from the ``conservation'' of
the generalized Komar charge Eq.~(\ref{eq:dK=0}). For the
asymptotically-flat, static, black-hole solutions of any $d=4$
ungauged supergravity this relation takes the form
\cite{Mitsios:2021zrn}

\begin{equation}
  \label{eq:Smarrrelation}
  \frac{1}{2}\left(M +\Phi^{M}_{\infty}\mathcal{Q}_{M} \right)
  = TS +\frac{1}{2}\Phi^{M}_{\mathcal{H}}\mathcal{Q}_{M}\,, 
\end{equation}

\noindent
where $\mathcal{Q}_{M}$ is a symplectic vector of electric and
magnetic charges and $\Phi^{M}$ is a symplectic vector of
electrostatic and magnetostatic potentials computed at infinity
($\Phi^{M}_{\infty}$) and at the event horizon
($\Phi^{M}_{\mathcal{H}}$).

Supersymmetric black holes must be extremal ($T=0$)
\cite{Gibbons:1984kp} and the Smarr relation
Eq.~(\ref{eq:Smarrrelation}) can be written in the form

\begin{equation}
  \label{eq:Smarrbound}
  M -\left(\Phi^{M}_{\infty}-\Phi^{M}_{\mathcal{H}}\right)\mathcal{Q}_{M}
  =
  0\,,
\end{equation}

\noindent
which is the form that a supersymmetry bound is expected to take if
$\Phi^{M}_{\infty}-\Phi^{M}_{\mathcal{H}}$ only depends on the moduli.

Since most proofs of the general supersymmetry bounds are based on
coordinate-dependent asymptotic expansions of general families of
solutions, the above arguments suggest that the generalized Komar
integrals could be used to derive rigorous, coordinate-independent
proofs of the supersymmetry bounds. IN this paper we are going to see
that this is indeed, possible, but we are going to prove, for the
cases considered, a more powerful result involving the generalized
Komar charge and not just its integral.

The first step in all the general studies of the supersymmetric
solutions of supergravity theories is the study of the differential
equation satisfied by the vector field $V$ that one can construct as a
bilinear of the Killing spinor $\epsilon$ they admit by hypothesis
$V^{\mu}\sim \bar{\epsilon}\gamma_{\mu}\epsilon$. We will call this
vector the \textit{supersymmetric vector}. The symmetric part of the
covariant derivative of this vector $\nabla_{(\mu}V_{\nu)}$ always
vanishes (\textit{i.e.}, $V$ is a Killing vector) and the
antisymmetric part has the form

\begin{equation}
  d\hat{V}+\cdots =0\,,
\end{equation}

\noindent
with $\hat{V}=V_{\mu}dx^{\mu}$. Thus, this equation involves the Hodge
dual of the standard Komar charge of pure gravity. In all the cases
that we have studied, the other terms in the equation are the Hodge
duals of the remaining terms in the generalized Komar charges of the
theories under consideration. Thus, we find that the generalized Komar
charges of these theories vanish identically for the supersymmetric
vector. The integral of the generalized Komar charge of the
supersymmetric vector at infinity, which, evidently, vanishes, gives
the left-hand side of the supersymmetry bound.

This result depends on the asymptotic normalization of the potentials
but it holds for the most natural choices which turn not to be those
for which these potentials vanish identically on the horizon, so that
the relation Eq.~(\ref{eq:Smarrbound}) only depends on their asymptotic
values.

The cases we are going to study are $\mathcal{N}=2,d=4$ supergravity
coupled to vector supermultiplets, in Section~\ref{sec-N2d4},
$\mathcal{N}=1,d=5$ (also called $\mathcal{N}=2$) supergravity coupled
to vector supermultiplets in Section~\ref{sec-N1d5} and pure
$\mathcal{N}=1,d=10$ supergravity in
Section~\ref{sec-n1d10}. Section~\ref{sec-discussion} contains a
discussion of the results and the problems encountered and future
directions of research.

\section{$\mathcal{N}=2,d=4$ supergravity coupled to vector supermultiplets}
\label{sec-N2d4}

One of the first, and general consequences of the KSEs is the
existence of a non-spacelike Killing vector $V$ whose dual 1-form
$\hat{V}$ satisfies the equation\footnote{This is Eq.~(3.24) of
  Ref.~\cite{Meessen:2006tu}.}

\begin{equation}
\label{eq:dV}
d\hat{V} - 4i (XT^{*\, -} -X^{*}T^{+})=0\,,
\end{equation}

\noindent
where $T$, the combination of scalars and vector field strengths that occurs
in the gravitini KSE is defined as

\begin{equation}
  T
  \equiv
  2i\mathcal{L}^{\Sigma}\Im{\rm m}\, \mathcal{N}_{\Sigma\Lambda}
  F^{\Lambda}\,.
\end{equation}

Using 

\begin{equation}
  \begin{aligned}
    \Im{\rm m}\, \mathcal{N}_{\Sigma\Lambda} F^{\Lambda\, +}
    & =
      \tfrac{1}{2}\Im{\rm m}\, \mathcal{N}_{\Sigma\Lambda} F^{\Lambda}
      +\tfrac{i}{2}  \Im{\rm m}\, \mathcal{N}_{\Sigma\Lambda} \star F^{\Lambda}
    \\
    & \\
    & =
      \tfrac{1}{2}\Im{\rm m}\, \mathcal{N}_{\Sigma\Lambda} F^{\Lambda}
      +\tfrac{i}{2}\left( F_{\Lambda}  -\Re{\rm e}\, \mathcal{N}_{\Sigma\Lambda} F^{\Lambda}\right)
    \\
    & \\
    & =
      \tfrac{1}{2}\left(\Im{\rm m} \, \mathcal{N}_{\Sigma\Lambda}
      -i \Re{\rm e}\, \mathcal{N}_{\Sigma\Lambda} \right) F^{\Lambda}
      +\tfrac{i}{2}F_{\Lambda}
    \\
    & \\
    & =
      -\tfrac{i}{2}\left(\mathcal{N}_{\Sigma\Lambda} F^{\Lambda} -F_{\Lambda}\right)\,,
  \end{aligned}
\end{equation}

\noindent
the self-dual part of $T$ can be rewritten as

\begin{equation}
  T^{+}
 =  
 \mathcal{L}^{\Lambda}F_{\Lambda}-\mathcal{M}_{\Lambda} F^{\Lambda}
 =
 \mathcal{V}^{M}F_{M}\,,
\end{equation}

\noindent
so that Eq.~(\ref{eq:dV}) takes the form

\begin{equation}
\label{eq:dV2}
d\hat{V} - 8|X|^{2} \mathcal{I}^{M}F_{M} = 0\,,
\end{equation}

\noindent
where we have defined real, K\"ahler-neutral, symplectic vectors

\begin{equation}
  \label{eq:RandIdef}
\mathcal{V^{M}}/X \equiv \mathcal{R}^{M}+i \mathcal{I}^{M}\,.  
\end{equation}

Now, let us take the Hodge dual of both sides of Eq.~(\ref{eq:dV2}):

\begin{equation}
\label{eq:dV3}
\star d\hat{V} - 8|X|^{2} \mathcal{I}^{M} \star F_{M} =0\,.
\end{equation}

The first term in this expression is proportional to the standard
Komar 2-form charge of General Relativity. When $V$ is a timelike
vector properly normalized at infinity, the integral of the first term
at spatial infinity (the standard Komar integral) gives $-M/2$. Since,
in the timelike case in adapted coordinates, $V=\sqrt{2}\partial_{t}$,
the integral of the first term at spacelike infinity will give
$-M/sqrt{2}$.

In general, when gravity is coupled to matter, the standard Komar
2-form charge is not closed on-shell and one has to use the
\textit{generalized Komar 2-form charge} which includes additional
terms. For generic 4-dimensional ungauged supergravity-like theories,
the generalized Komar 2-form charge for any Killing vector field $V$
was found in Refs.~\cite{Mitsios:2021zrn,Ballesteros:2023iqb} and in
the conventions used in that reference it takes the form

\begin{equation}
  \mathbf{K}[V]
  =
  \frac{1}{16\pi G_{N}^{(4)}}\left[ -\star dV +\tfrac{1}{2}P_{V}{}^{M}F_{M}\right]\,.
\end{equation}

\noindent
where $P_{V}$ satisfies the momentum map equation

\begin{equation}
\imath_{V}F^{M}+dP_{V}{}^{M} =0\,.  
\end{equation}

The gauge fields used in Ref.~\cite{Meessen:2006tu}, which we are
using here, are $2\sqrt{2}$ times those used in
Refs.~\cite{Mitsios:2021zrn,Ballesteros:2023iqb}. Thus, the
generalized Komar charge takes the form 

\begin{equation}
  \label{eq:Komarcharge}
  \mathbf{K}[V]
  =
  -\frac{1}{16\pi G_{N}^{(4)}}\left\{\star d\hat{V}
    +4 E^{M}F_{M}\right\}\,,  
\end{equation}

\noindent
where $E^{M}$ is defined, as in Eq.~(4.25) of Ref.~\cite{Meessen:2006tu}, by
the equation

\begin{equation}
  \label{eq:electrictaticpotentialsdef}
d E^{M} = \imath_{V}F^{M}\,.  
\end{equation}

In Ref.~\cite{Meessen:2006tu} it was also found that unbroken supersymmetry
implies that

\begin{equation}
  \label{eq:electristaticpotentialssolution}
  E^{M} = 2|X|^{2} \mathcal{R}^{M}\,, 
\end{equation}

\noindent
and we can rewrite the generalized Komar charge Eq.~(\ref{eq:Komarcharge}) in
the form

\begin{equation}
  \label{eq:Komarcharge2}
  \mathbf{K}[V]
  =
  -\frac{1}{16\pi G_{N}^{(4)}}\left\{\star d\hat{V}
  +8|X|^{2} \mathcal{R}^{M}F_{M}\right\}\,.  
\end{equation}

This expression is very similar to the left-hand side of
Eq.~(\ref{eq:dV3}), required by supersymmetry. In order to bring the
left-hand side of Eq.~(\ref{eq:dV3}) to a form closer to that of
Eq.~(\ref{eq:Komarcharge2}), we use the \textit{twisted self-duality
  constraint} (see, for instance Ref.~\cite{Ortin:2015hya})

\begin{equation}
  \label{eq:twisted}
  \star F_{M} = -\mathcal{M}_{MN}(\mathcal{N})F^{N}\,,   
\end{equation}

\noindent
to eliminate the Hodge duals of the field strengths in terms of the field
strengths themselves getting

\begin{equation}
\label{eq:dV4}
\star d\hat{V} +8|X|^{2} \mathcal{I}^{M} \mathcal{M}_{MN}(\mathcal{N})F^{N}  =0\,.
\end{equation}

In the above expressions, the symmetric symplectic matrix
$\mathcal{M}_{MN}(\mathcal{A})$

\begin{equation}
\mathcal{M}(\mathcal{A}) 
\equiv 
\left(
  \begin{array}{cc}
\Im\mathfrak{m}\, \mathcal{A}_{\Lambda\Sigma} +
\Re\mathfrak{e}\, \mathcal{A}_{\Lambda\Omega}\, 
\Im\mathfrak{m}\, \mathcal{A}^{-1|\, \Omega\Gamma}\,
\Re\mathfrak{e}\, \mathcal{A}_{\Gamma\Sigma} 
& 
\hspace{.5cm}
-\Re\mathfrak{e}\, \mathcal{A}_{\Lambda\Omega}\,
\Im\mathfrak{m}\, \mathcal{A}^{-1 |\, \Omega\Sigma}
\\
\\
-
\Im\mathfrak{m}\, \mathcal{A}^{-1 |\, \Lambda\Omega}\,
\Re\mathfrak{e}\, \mathcal{A}_{\Omega\Sigma}
&
\Im\mathfrak{m}\, \mathcal{A}^{-1|\, \Lambda\Sigma}
\end{array}
\right),
\end{equation}

\noindent
can be associated with any symmetric complex matrix
$\mathcal{A}_{\Lambda\Sigma}$ with a non-degenerate imaginary part such as the
period matrix $\mathcal{N}_{\Lambda\Sigma}$ in our case. As explained in
Ref.~\cite{Meessen:2011aa}, if 

\begin{equation}
\mathcal{M}_{\Lambda} = \mathcal{N}_{\Lambda\Sigma}\mathcal{L}^{\Sigma}\,, 
\end{equation}

\noindent
(as it is the case, by definition of the period matrix) and we define the
real, K\"ahler-neutral, symplectic vectors $\mathcal{R}^{\Lambda}$ and
$\mathcal{I}^{\Lambda}$ by Eq.~(\ref{eq:RandIdef}), then

\begin{subequations}
  \begin{align}
    \label{eq:RMI}
    \mathcal{R}_{M}
    & =
      -\mathcal{M}_{MN}(\mathcal{N}) \mathcal{I}^{N}\,,
    \\
    & \nonumber \\
    \label{eq:dRMdI}
    d\mathcal{R}_{M}
    & =
      -\mathcal{M}_{MN}(\mathcal{N}) d\mathcal{I}^{N}\,.
  \end{align}
\end{subequations}

Using the first of these properties in Eq.~(\ref{eq:dV4}) we find that
this equation, implied by unbroken supersymmetry, is fully equivalent
to the vanishing of the generalized Komar charge written in
Eq.~(\ref{eq:Komarcharge2})

\begin{equation}
  \label{eq:KV=0}
\mathbf{K}[V] = 0\,.  
\end{equation}

This equation is satisfied both for timelike and null vectors, although in a
rather trivial way in the second case because the null Killing vectors are
covariantly constant and also $\imath_{V}T^{+}=0$. Thus, we will focus on the
timelike case in what follows.

\subsection{Consequences for the charges}
\label{sec-consequencesd4}

When using Eq.~(\ref{eq:KV=0}) we must take into account the fact that the
Komar charge is defined only up to total derivatives. In particular, the above
equation is true only when the potentials $E^{M}$ defined by
Eq.~(\ref{eq:electrictaticpotentialsdef}) are identified as in
Eq.~(\ref{eq:electristaticpotentialssolution}) with no additive integration
constants whatsoever. Adding a symplectic vector of constants $C^{M}$ to the
potentials shifts $\mathbf{K}[V]$ by the closed, but non-vanishing 2-form
$C^{M}F_{M}$. Thus, it is important to understand what were the boundary
conditions of the potentials that made the generalized Komar charge
identically vanish.

In timelike solutions $2|X|^{2} =g_{tt}$ which should be normalized to $1$ at
infinity and should vanish on the horizon of a black-hole spacetime. This means
that $\star d\hat{V}$ will also vanish on that horizon and, if it has finite
area, it should have vanishing temperature, characterizing the black hole as
an extremal one.

At infinity, we expect the integral of the term $\mathcal{R}^{M}F_{M}$ to be
related to the asymptotic value of the central charge of the theory

\begin{equation}
  \mathcal{Z}(Z,\mathcal{Q}) \equiv \mathcal{V}_{M}\mathcal{Q}^{M}
  = e^{i\alpha} |X| \left[\left(\mathcal{R}^{M}\mathcal{Q}_{M}\right)
    +i\left(\mathcal{I}^{M}\mathcal{Q}_{M}\right)\right]\,,  
\end{equation}

\noindent
where the symplectic vector of electric and magnetic charges $\mathcal{Q}^{M}$
is defined by

\begin{equation}
  \mathcal{Q}^{M}
  \equiv 
  \frac{1}{2\pi G_{N}^{(4)}} \int_{S^{2}_{\infty}}F^{M}\,.
\end{equation}

\noindent
and where $\alpha$ is a local phase defined up to a constant that we
can define so that, asymptotically,

\begin{equation}
  |\mathcal{Z}_{\infty}(Z,\mathcal{Q})|
  =
  \frac{1}{\sqrt{2}} \left(\mathcal{R}^{M}_{\infty}\mathcal{Q}_{M}\right)\,.  
\end{equation}

Then, the integral of the generalized Komar charge associated to the
Killing vector $V= \sqrt{2}\partial_{t}$ at spatial infinity gives

\begin{equation}
  \int_{S^{2}_{\infty}}\mathbf{K}[V]
  =
  \sqrt{2} \frac{M}{2} - \frac{1}{2}R^{M}_{\infty}\mathcal{Q}_{M}\,,   
\end{equation}

\noindent
and supersymmetry implies the $\mathcal{N}=2$ supersymmetry bound

\begin{equation}
M = |\mathcal{Z}_{\infty}(Z,\mathcal{Q})|\,.  
\end{equation}

\section{$\mathcal{N}=1,d=5$ supergravity coupled to vector supermultiplets}
\label{sec-N1d5}

Again, our starting point is the equation for the antisymmetric part
of the derivative of the supersymmetric Killing vector $V$ (the
Killing vector constructed as a bilinear of the Killing
spinor)\footnote{This is Eq.~(4.6) of Ref.~\cite{Bellorin:2006yr},
  whose conventions we use here, up to the replacement of $V$ by
  $\hat{V}$ to denote the 1-form dual to the Killing vector $V$.}:

\begin{equation}
  \label{eq:dV-d5}
  d\hat{V}
  =
  -\tfrac{2}{\sqrt{3}}fh_{I}F^{I}
-\tfrac{1}{\sqrt{3}} h_{I} \star\left(F^{I}\wedge \hat{V}\right)\,.
\end{equation}

\noindent
$F^{I}$ are the $n+1$ 2-form gauge field strengths
$I,J,K=0,1,\ldots,n+1$, $h^{I}(\phi)$ are functions of the $n$ scalars
$\phi^{x}$, $x,y,z=1,\ldots,n$, satisfying the constraint

\begin{equation}
  \label{eq:constraint}
C_{IJK}h^{I}h^{J}h^{K}=1\,,
\end{equation}

\noindent
and

\begin{equation}
h_{I}\equiv C_{IJK} h^{J}h^{K}\, ,\,\,\,\, \Rightarrow h_{I}h^{I}=1\,.
\end{equation}

The metric

\begin{equation}
a_{IJ}=-2C_{IJK}h^{K} +3h_{I}h_{J}\,,  
\end{equation}

\noindent
and its inverse $a^{IJ}$ can be used to raise and lower $I,J,K,\ldots$
indices. We will mention more objects and properties of the real
special geometry governing these theories as the need arises.

The function $f$ that occurs in Eq.~(\ref{eq:dV-d5}) is related to the supersymmetric Killing vector $V$ by

\begin{equation}
V^{2} = f^{2} \geq 0\,.  
\end{equation}

Thus, as in the 4-dimensional case, the Killing vector can be timelike
or null, but, as different from the 4-dimensional case, when the
Killing vector is null, it is not covariantly constant and the
vanishing of the Komar charge is not trivial. We have to consider each
of these two cases separately. 

\subsection{The timelike case}

In this case $V=\partial_{t}$ is the timelike Killing vector in
adapted coordinates in which the metric takes the form

\begin{subequations}
  \begin{align}
    \label{conforma-stationary}
    ds^{2}
    & =
    f^{2}\left(dt+\omega\right)^{2}
      -f^{-1}ds_{(4)}^{2}\,,
    \\
    & \nonumber \\
    ds_{(4)}^{2}
    & \equiv
      h_{\underline{m}\underline{n}} dx^{m}dx^{n}\,,
  \end{align}
\end{subequations}

\noindent
so that

\begin{equation}
\hat{V} = f^{2}\left(dt+\omega\right)\,,  
\end{equation}

The Hodge dual of Eq.~(\ref{eq:dV-d5}) is

\begin{equation}
  \label{eq:dV2-d5}
  \star d\hat{V}
    +\tfrac{2}{\sqrt{3}}fh_{I} \star F^{I}
    +\tfrac{1}{\sqrt{3}} h_{I} F^{I}\wedge \hat{V}
    =
    0\,.
\end{equation}

The 2-form field strengths of the supersymmetric solutions
are\footnote{This is Eq.~(4.20) of Ref.~\cite{Bellorin:2006yr}.}

\begin{equation}
\label{eq:FI}
F^{I} = -\sqrt{3}\left\{d\left[fh^{I}\left(dt+\omega\right)\right]
+\Theta^{I}\right\}\, , 
\end{equation}

\noindent
where the $\Theta^{I}$s are selfdual 2-forms in the 4-dimensional
Euclidean \textit{base space} defined by the metric
$h_{\underline{m}\underline{n}}$. They are related to the self-dual
part of 1-form $fd\omega$ in the metric by\footnote{These are
  Eqs.~(4.18) and (4.21) of Ref.~\cite{Bellorin:2006yr}.}

\begin{equation}
\label{remanentconstraint}
h_{I}\Theta^{I}
=
-\tfrac{2}{3} (fd\omega)^{+}\,.
\end{equation}

\noindent
Thus, they are orthogonal to $V$ and, then,

\begin{equation}
  \imath_{V}F^{I}
  =
  -\sqrt{3} \imath_{V}d\left[fh^{I}\left(dt+\omega\right)\right]
  =
  \sqrt{3} d\imath_{V}\left[fh^{I}\left(dt+\omega\right)\right]
  =
   d\left(\sqrt{3}fh^{I}\right)\,,
\end{equation}

\noindent
which implies that the electric momentum maps associated to the
Killing vector $V$, defined in
Eq.~(\ref{eq:electricmomentummapequations1}), are, up to additive
constants,

\begin{equation}
  \label{eq:electricmomentummapsd5}
  P_{V}{}^{I}
  =
  -\sqrt{3}fh^{I}\,,
\end{equation}

\noindent
and Eq.~(\ref{eq:dV2-d5}) can be put in the form

\begin{equation}
  \label{eq:dV3-d5}
  \star d\hat{V}
    -\tfrac{2}{3} P_{V}{}^{I}a_{IJ}\star F^{J}
    +\tfrac{1}{\sqrt{3}} h_{I} F^{I}\wedge \hat{V}
    =
    0\,.
\end{equation}

Let us now compute the magnetic momentum maps $\tilde{P}_{V\, I}$ defined in
Eq.~(\ref{eq:magneticmomentummapequation}). First, we compute

\begin{equation}
  \begin{aligned}
    \imath_{V}\star F^{I}
    & =
      -\sqrt{3}f\left[\left(\delta^{I}{}_{J}-\tfrac{3}{2}h^{I}h_{J}\right)\Theta^{J}
      -h^{I}\left(fd\omega\right)^{-}\right]\,,
  \end{aligned}
\end{equation}

\noindent
and

\begin{equation}
  \imath_{V}\left(a_{IJ}\star F^{J}\right)
  =
      -\sqrt{3}f\left[\left(a_{IJ}-\tfrac{3}{2}h_{I}h_{J}\right)\Theta^{J}
      -h_{I}\left(fd\omega\right)^{-}\right]\,.
\end{equation}

Then, we  compute 

\begin{equation}
  \begin{aligned}
    -\tfrac{2}{\sqrt{3}}C_{IJK}P_{V}{}^{J}F^{K}
    & =
    d\left(-\sqrt{3}h_{I}\hat{V}\right)
      -\sqrt{3}f^{2}h_{I}d\omega
      -2\sqrt{3}fC_{IJK}h^{J}\Theta^{K}\,.
  \end{aligned}
\end{equation}

The sum of these two results gives

\begin{equation}
  \begin{aligned}
  \imath_{V}\left(a_{IJ}\star F^{J}\right)
      -\tfrac{2}{\sqrt{3}}C_{IJK}P_{V}{}^{J}F^{K}
  & =
      d\left(-\sqrt{3}h_{I}\hat{V}\right)\,,
  \end{aligned}
\end{equation}

\noindent
which implies that

\begin{equation}
    \label{eq:magneticmomentummapsd5}
  \tilde{P}_{V\, I}
  =
  \sqrt{3}h_{I}\hat{V}\,,
  \hspace{1cm}
  \mathfrak{h}^{(2)}{}_{I}
  =
  0\,.
\end{equation}

This is the result obtained by Gibbons and Warner in
Ref.~\cite{Gibbons:2013tqa} and it allows us to rewrite
Eq.~(\ref{eq:dV3-d5}) in the form

\begin{equation}
  \label{eq:dV4-d5}
  \mathbf{K}[V]
    =
    0\,,
\end{equation}

\noindent
where

\begin{equation}
  \label{eq:KVd5}
  \mathbf{K}[V]
  \equiv
  \star d\hat{V}
    -\tfrac{2}{3} P_{V}{}^{I}a_{IJ}\star F^{J}
    +\tfrac{1}{3} \tilde{P}_{V\, I}\wedge F^{I}\,,  
\end{equation}

\noindent
is the generalized Komar 3-form charge of this theory.

\subsection{The null case}

In this case the function $f$ in Eq.~(\ref{eq:dV-d5}) vanishes
identically. Calling the null supersymmetric Killing vector $l$,
Eq.~(\ref{eq:dV-d5}) takes the form

\begin{equation}
  \label{eq:dV-d5-null}
  d\hat{l}
  =
-\tfrac{1}{\sqrt{3}} h_{I} \star\left(F^{I}\wedge \hat{l}\,\right)\,.
\end{equation}

Introducing two coordinates $u,v$ such that $l= \partial_{v}$ and
$\hat{l} = fdu$, the metric and the gauge field strengths can be
written in the form\footnote{These are Eqs.~(4.88) and (4.101) of
  Ref.~\cite{Bellorin:2006yr}. Notice that the function $f$ used in
  the null case has nothing to do with the original function $f$
  defined as the norm of the supersymmetric Killing vector, which
  vanishes in this case.}

\begin{subequations}
  \begin{align}
    ds^{2}
    & =
      2f du(dv +Hdu +\omega) -f^{-2}\gamma_{\underline{r} \underline{s}}dx^{r}dx^{s}\,,
    \\
    & \nonumber \\
    \label{eq:FInull}
    F^{I}
    & = 
[\tfrac{1}{\sqrt{3}} f^{2}h^{I}\hat{\star}d\omega -\psi^{I}]\wedge du
+\sqrt{3}\hat{\star} \hat{d}(h^{I}/f)\,,
  \end{align}
\end{subequations}

\noindent
where $r,s,t=1,2,3$, $\hat{\star}$ and $\hat{d}$ stand for the Hodge
star and exterior derivative in the 3-dimensional space with
Ricci-flat metric $\gamma_{\underline{r} \underline{s}}$, the 1-forms
$\psi^{I}$ satisfy $h_{I}\psi^{I}=0$ and all the objects that appear
in these expressions are $v$-independent.

From Eq.~(\ref{eq:FInull}) we find that\footnote{This is Eq.~(4.94)
of Ref.~\cite{Bellorin:2006yr}.} 

\begin{equation}
  \imath_{l}F^{I}
  =
  0\,,
\end{equation}

\noindent
which implies that we can take the electric momentum maps as
constants. For simplicity we choose $P_{l}{}^{I}=0$.

We also find that 

\begin{equation}
  \imath_{l} \star  F^{I}
  =
  \sqrt{3}f^{2} d (h^{I}/f)\wedge du\,,
\end{equation}

\noindent
so

\begin{equation}
  \begin{aligned}
    \imath_{l}\left(a_{IJ} \star  F^{J}\right)
    & =
    \sqrt{3}fa_{IJ}dh^{J}\wedge du
      -\sqrt{3} a_{IJ}h^{J}df\wedge du
    \\
    & \\
    & =
    -\sqrt{3}dh_{I}\wedge fdu
      -\sqrt{3} h_{I}df\wedge du
    \\
    & \\
    & =
    d\left(-\sqrt{3}h_{I}\hat{l}\right)\,,
  \end{aligned}
\end{equation}

\noindent
and, owing to the vanishing of the electric momentum maps, we find
that the magnetic ones are given by

\begin{equation}
  \label{eq:d5nullmagneticmomentummaps}
  \tilde{P}_{l\, I}
  =
  \sqrt{3}h_{I}\hat{l}\,,
\end{equation}

\noindent
and we can rewrite the Hodge dual of the supersymmetric vector
equation (\ref{eq:dV-d5-null}) as

\begin{equation}
  \mathbf{K}[l]
  =
  0\,.
\end{equation}

\subsection{Consequences for the charges}
\label{sec-consequencesd5}

The Komar charge vanishes for the supersymmetric Killing vector with
the choice of momentum maps that we have made:
Eqs.~(\ref{eq:electricmomentummapsd5}) and
(\ref{eq:magneticmomentummapsd5}). Let us explore the implications as
in the 4-dimensional case, including in the definition of the Komar
charge the overall normalization factor $(16\pi G_{N}^{(5)})^{-1}$.

Consider an asymptotically-flat solution whose gauge fields are purely
electric $A^{I}=A^{I}{}_{t}dt$. The pullback of the last term of the
generalized Komar charge Eq.~(\ref{eq:KVd5}) over the 3-sphere at
spatial infinity vanishes identically. The integral of the first term,
which is nothing but the standard Komar integral for $V=\partial_{t}$
gives $2M/3$ while the second gives

\begin{equation}
  \begin{aligned}
    \frac{1}{16\pi G_{N}^{(5)}}
    \int_{S^{3}_{\infty}}\left[ -\tfrac{2}{3} P_{V}{}^{I}a_{IJ}\star F^{J}  \right]
    & =
    \tfrac{2}{\sqrt{3}} \frac{1}{16\pi G_{N}^{(5)}}
      \int_{S^{3}_{\infty}}\left[  fh^{I}a_{IJ}\star F^{J}  \right]
    \\
    & \\
    & =
    \tfrac{2}{\sqrt{3}}h^{I}_{\infty} \frac{1}{16\pi G_{N}^{(5)}}
      \int_{S^{3}_{\infty}}\left[  a_{IJ}\star F^{J}  \right]
    \\
    & \\
    & =
    -\tfrac{2}{\sqrt{3}} h^{I}_{\infty}q_{I}
    \\
    & \\
    & =
    -\tfrac{2}{\sqrt{3}} \mathcal{Z}_{\infty}(\phi_{\infty},q)\,,
  \end{aligned}
\end{equation}

\noindent
where $\phi^{x}_{\infty}$ are the asymptotic values of the
scalars\footnote{The scalars can be defined in terms of the functions
  $h^{I}$, $I=0,x$, $x=1,\cdots, n$ by

  \begin{equation}
  \phi^{x} = h^{x}/h^{0}\,.  
\end{equation}

This equation and Eq.~(\ref{eq:constraint}) can be used to write the
$n+1$  $h^{I}$ as functions of the $n$ scalars.

} and
$\mathcal{Z}_{\infty}(\phi_{\infty},q)$ is the central charge of the
theory. Thus, the vanishing of the Komar charge implies the supersymmetry bound

\begin{equation}
  M = \sqrt{3}\mathcal{Z}_{\infty}(\phi_{\infty},q)\,.
\end{equation}

Solutions with purely magnetic gauge fields are solutions with purely
electric $B_{I}$s: strings. They fall within the class of
supersymmetric solutions for which the supersymmetric Killing vector
$V$ is null.

These solutions have an additional translational isometry along the
string and the standard Komar integral diverges unless the string is
wrapped around a compact dimension. On the other hand, if we want to
derive a supersymmetry bound using the vanishing of the generalized
Komar charge, we have to take into account that this only happens for
the supersymmetric Killing vector, which is null and the physical
interpretation of the integrals of the different terms that occur in
the generalized Komar charge is unclear. We expect a bound relating
the string tension to the magnetic charges but the integral seems to
give momentum.

This problem arises because the generalized Komar charge is a
$(d-2)$-form and it is meant to compute the charges of point-like
objects with a codimension 1 space transverse to their worldvolumes
(worldlines). A different tool is needed to deal with extended objects
with lower-dimensional transverse spaces \cite{kn:BCO}.

\section{Pure $\mathcal{N}=1,d=10$ supergravity}
\label{sec-n1d10}

Pure $\mathcal{N}=1,d=10$ supergravity is the low-energy effective
field theory of the heterotic string to lowest order in $\alpha'$. The
Noether--Wald charge associated to any vector field
$\xi=\xi^{\mu}\partial_{\mu}$ of this theory has been calculated in
Refs.~\cite{Elgood:2020mdx,Elgood:2020nls} and, in the string frame,
takes the form\footnote{This is Eq.~(3.65) of
  Ref.~\cite{Elgood:2020nls}.}

\begin{equation}
  \mathbf{Q}[\xi]
  = 
  e^{-2\phi}\left[\star d\hat{\xi}
    +\star H \wedge P_{\xi}\right]
  -2 \left(\star de^{-2\phi}\wedge \hat{\xi}\right)\,,
\end{equation}

\noindent
where $\hat{\xi}=\xi_{\mu}dx^{\mu}$, $H=dB$ is the 3-form field
strength of the Kalb--Ramond 2-form $B$ and $P_{\xi}$ is the 1-form
momentum map associated to $B$. It is assumed to satisfy the momentum
map equation\footnote{There are no closed but not exact 2-forms in
  this equations because, as pointed out in
  Ref.~\cite{Barbagallo:2025qdy}, this would imply that $B$ is not
  invariant up to, at most, a gauge transformation, under the GCT
  generated by $k$.}

\begin{equation}
  \label{eq:momentummapequationH}
\imath_{k}H +dP_{k} = 0\,,  
\end{equation}

\noindent
when $\xi=k$, a vector field generating a GCT that leaves invariant all the
field of the theory which implies that, in particular, it is a Killing vector
of the metric.

Finally, we are ignoring for the moment an overall factor

\begin{equation}
\frac{g_{s}^{2}}{16\pi G_{N}^{(10)}}\,,
\end{equation}

\noindent
which we will add to the final results in order to keep in the intermediate
expressions as simple as possible.

Since the action of this theory is exactly gauge invariant, for such a vector,
on shell

\begin{equation}
\mathbf{Q}[k]
    \doteq
\imath_{k}\mathbf{L}\,,  
\end{equation}

\noindent
and, following Ref.~\cite{Ortin:2024mmg}, in order to construct the
generalized Komar charge we first have to rewrite the right-hand side of this
expression (after evaluating $\mathbf{L}$ on-shell) as a total derivative

\begin{equation}
  \imath_{k}\mathbf{L}
  =
  d\boldsymbol{\omega}_{k}\,,
\end{equation}

\noindent
and define the Komar charge as

\begin{equation}
  \mathbf{K}[k]
  \equiv
  -\mathbf{Q}[k]
  +\boldsymbol{\omega}_{k}\,.  
\end{equation}

The equation of motion of the dilaton is

\begin{equation}
  \mathbf{E}_{\phi}
  =
  -4 d \star d e^{-2\phi} -2\mathbf{L}\,,
\end{equation}

\noindent
where

\begin{equation}
  \mathbf{L}
  =
  e^{-2\phi}\left\{-\star (e^{a}\wedge e^{b})\wedge R_{ab} -4 d\phi\wedge
    \star d\phi +\tfrac{1}{2}H\wedge \star H \right\}\,,
\end{equation}

\noindent
is the Lagrangian of the theory so, on-shell, the Lagrangian is given by

\begin{equation}
  \mathbf{L}
  \doteq
  -2  d \star d e^{-2\phi}\,.
\end{equation}

Since, by assumption, the GCT generated by $k$ leaves all the fields of the
solution considered invariant, we find that 

\begin{equation}
  \begin{aligned}
    \imath_{k}\mathbf{L}
    & \doteq
      d\star\left( -2d e^{-2\phi}\wedge \hat{k}\right)
 \equiv
    d\boldsymbol{\omega}_{k}\,,
  \end{aligned}
\end{equation}

\noindent
and the Komar charge is given by the deceptively simple expression

\begin{equation}
  \mathbf{K}[k]
  =
  \frac{g_{s}^{2}}{16\pi G_{N}^{(10)}}
  e^{-2\phi}\left[\star d\hat{k}
    +\star H \wedge P_{k}\right]\,.
\end{equation}

All the supersymmetric configurations of this theory admit a null
Killing vector, $\ell$, with respect to which, the following equation
is satisfied\footnote{This is Eq.~(3.9) of
  Ref.~\cite{Fontanella:2019avn}.}

\begin{equation}
\imath_{\ell}H -d\hat{\ell} = 0\,.  
\end{equation}

Comparing this equation with Eq.~(\ref{eq:momentummapequationH}), we conclude
that the momentum map 1-form associated to the Killing vector $\ell$ is
just $-\hat{\ell}$

\begin{equation}
  P_{\ell}  = -\hat{\ell}\,,
\end{equation}

\noindent
and we can write

\begin{equation}
  \mathbf{K}[\ell]
  =
  \frac{g_{s}^{2}}{16\pi G_{N}^{(10)}}
  e^{-2\phi}\left[\star d\hat{\ell}
    -\star H \wedge \hat{\ell}\right]
  =
  \frac{g_{s}^{2}}{16\pi G_{N}^{(10)}}
  e^{-2\phi}\star\left[ d\hat{\ell}
    -\imath_{\ell}H \right]
  =
  0\,.
\end{equation}

Notice that in the 5-dimensional null case the magnetic momentum maps
1-forms, given in Eq.~(\ref{eq:d5nullmagneticmomentummaps}), are also
proportional to the supersymmetric Killing vector.

\subsection{Consequences for the charges}
\label{sec-consequencesd10}

Again, the natural solutions of this theory for which we could derive
a supersymmetry bound are strings and the inadequacy of the
generalized Komar $(d-2)$-form charge to give meaningful results
discussed in the previous section prevents us from doing so.  Clearly,
to this end, a not-yet-available generalization of the Komar charge
for extended objects is necessary \cite{kn:BCO}.

\section{Conclusions}
\label{sec-discussion}

While the results that we have obtained are reasonable or even
expected on general grounds, we have had to derive them case by case
and we do not know any general proof or construction applicable to
every theory of supergravity.  To the best of our knowledge, the
generalized Komar $(d-2)$-form charge is not related to a
supersymmetry transformation that would vanish for Killing spinors.
Finding a construction that leads to this on-shell closed
$(d-2)$-form charge is a challenging but very interesting problem.

Meanwhile, it is desirable to extend these results to other theories
and also to study the effect of the first $\alpha'$ corrections in the
effective action of the heterotic superstring, which is possible
because they have been found using supersymmetry as a guiding
principle in Ref.~\cite{Bergshoeff:1989de}. Work in this direction is
already under way \cite{kn:BO}.

In the cases in which the supersymmetric Killing vector is null and
the typical solutions are extended objects characterized by additional
translational symmetries, we have been unable to derive supersymmetry
bounds because the naive Komar integrals diverge. A further
generalization of the Komar charge, adapted to objects with
$(p+1)$-dimensional worldvolumes and $(d-p-1)$ transverse spaces (a
$(d-p-2)$-form) is clearly needed. The usual techniques associated to
standard symmetries, generated by single Killing vectors, cannot
directly provide it, but other possibilities are currently being
explored \cite{kn:BCO}.

\section*{Acknowledgments}

The work of TO has been supported in part by the MCI, AEI, FEDER
(UE) grants PID2021-125700NB-C21 and PID2024-155685NB-C21 (``Gravity,
Supergravity and Superstrings'' (GRASS)) and IFT Centro de Excelencia Severo
Ochoa CEX2020-001007-S. TO wishes to thank M.M.~Fern\'andez for her
permanent support.

\appendix

\section{The electric and magnetic momentum maps in $\mathcal{N}=1,d=5$
  supergravity}
\label{sec-momentummap}

Assuming that all the fields are left invariant by the GCT generated by the
Killing vector $V$ and using the Bianchi identities $dF^{I}=0$, we find that 

\begin{equation}
  0
  =
  \delta_{V}F^{I}
  =
  -\mathcal{L}_{V}F^{I}
  =
  -d\imath_{V}F^{I}\,,
\end{equation}

\noindent
which implies

\begin{equation}
  \label{eq:electricmomentummapequations1}
  \imath_{V}F^{I}
  =
  -dP_{V}{}^{I}\,,
\end{equation}

\noindent
where the functions $P_{V}{}^{I}$ defined by this equation are the
\textit{electric momentum maps}. In principle, one could add a closed
non-exact 1-form $\mathfrak{h}^{(1)\, I}$ to the right-hand side of
this equation if the topology of the solution admits it, but we are
going to show that it is incompatible with the invariance of the
gauge fields $A^{I}$.

Let us consider the invariance of the gauge fields $A^{I}$ under the same
GCT. This invariance can only be defined up to \textit{compensating} or
\textit{induced} gauge transformations with parameters $\Lambda_{V}{}^{I}$
and, therefore, we have

\begin{equation}
  0
  =
  \delta_{V}A^{I}
  =
  -\mathcal{L}_{V}A^{I}+d\Lambda_{V}{}^{I}
  =
  -\imath_{V}F^{I} +d\left(\Lambda_{V}{}^{I}-\imath_{V}A^{I} \right)\,.
\end{equation}

\noindent
Using Eq.~(\ref{eq:electricmomentummapequations1}), we
get\footnote{Had we included 1-forms $\mathfrak{h}^{(1)\, I}$ in the
  right-hand side of Eq.~(\ref{eq:electricmomentummapequations1}), we
  would have found an additional term $-\mathfrak{h}^{(1)\, I}$ in the
  right-hand side of the above equation. This is inconsistent with the
  assumption $ \delta_{V}A^{I}=$. However, we could have asked
  invariance of the gauge fields up to a global higher-form-type
  symmetry of the form

\begin{equation}
  \delta_{\eta}A^{I}
  =
  \eta \mathfrak{h}^{(1)\, I}\,,
\end{equation}

\noindent
that is

\begin{equation}
  \delta_{V}A^{I}
  =
  \delta_{\eta}A^{I}\,. 
\end{equation}

\noindent
This, weaker, symmetry condition, is compatible with
Eq.~(\ref{eq:electricmomentummapequations1}) for $\eta=-1$. We will,
however, not consider this possibility here.}

\begin{equation}
  0
  =
  d\left[\Lambda_{V}{}^{I}-\left(\imath_{V}A^{I}-P_{V}{}^{I}\right) \right]\,,
  -\mathfrak{h}^{(1)\, I}\,.
\end{equation}

\noindent
which requires the following identification of the gauge parameters

\begin{equation}
  \Lambda_{V}{}^{I}
  =
  \imath_{V}A^{I}-P_{V}{}^{I}\,.  
\end{equation}

The equations of motion of the Abelian gauged fields in ungauged
$\mathcal{N}=1,d=5$ supergravity coupled to vector supermultiplets are

\begin{equation}
    \mathbf{E}_{I}
     =
      -d\left(a_{IJ}\star F^{J} \right)
+\tfrac{1}{3^{1/2}}C_{IJK}F^{J}\wedge F^{K}\,.
\end{equation}

Assuming again that all the fields are left invariant by the GCT
generated by the Killing vector $V$ and using
Eq.~(\ref{eq:electricmomentummapequations1}) we find

\begin{equation}
  \begin{aligned}
    \imath_{V}\mathbf{E}_{I}
     & =
      d\left[\imath_{V}\left(a_{IJ}\star F^{J} \right)
      -\tfrac{2}{3^{1/2}}C_{IJK}P_{V}{}^{J} F^{K}\right]\,.
  \end{aligned}
\end{equation}

We conclude that

\begin{equation}
  \label{eq:magneticmomentummapequation}
\imath_{V}\left(a_{IJ}\star F^{J} \right)
-\tfrac{2}{3^{1/2}}C_{IJK}P_{V}{}^{J}F^{K}
\doteq
-d\tilde{P}_{V\, I}
-\mathfrak{h}^{(2)}_{V\, I}\,,
\end{equation}

\noindent
where, by assumption, the \textit{magnetic momentum map}
$\tilde{P}_{V\, I}$, defined by this equation, is globally defined and
$\mathfrak{h}^{(2)}_{V\, I}$ is closed but not exact
\cite{Gibbons:2013tqa} (see also Ref.~\cite{Barbagallo:2025qdy}). This
is the \textit{magnetic momentum map equation}.

It is interesting to analyze this equation from the point of view of the
2-forms $B_{I}$ dual to the gauge fields $A^{I}$. In order to determine their
gauge transformations and gauge field strengths, we first rewrite the
equations of motion of the gauge fields in the form

\begin{equation}
    \mathbf{E}_{I}
     =
      -d\left(a_{IJ}\star F^{J} 
-\tfrac{1}{3^{1/2}}C_{IJK}A^{J}\wedge F^{K}\right)\,.
\end{equation}

\noindent
that can be locally solved by 

\begin{equation}
a_{IJ}\star F^{J} 
-\tfrac{1}{3^{1/2}}C_{IJK}A^{J}\wedge F^{K}
\equiv
dB_{I}\,,
\,\,\,\,\,
\Rightarrow
\,\,\,\,\,
a_{IJ}\star F^{J}
=
dB_{I}+\tfrac{1}{3^{1/2}}C_{IJK}A^{J}\wedge F^{K}
\equiv
H_{I}\,.
\end{equation}

$H_{I}$ are the gauge-invariant 3-form field strengths of the 2-forms
$B_{I}$, which implies that these and the gauge fields transform as

\begin{subequations}
  \begin{align}
      \delta_{\Lambda}A^{I}
    & =
      d\Lambda^{(0)\, I}\,,
    \\
    & \nonumber \\
  \delta_{\Lambda}B_{I}
  & =
  d\Lambda^{(1)}{}_{I}
  -\tfrac{1}{3^{1/2}}C_{IJK}\Lambda^{(0)\, J} F^{K}\,.
  \end{align}
\end{subequations}

Let us now consider the invariance of the 2-forms under the GCT generated by
$V$:

\begin{equation}
  \begin{aligned}
  0
  & =
    \delta_{V}B_{I}
    \\
    & \\
  & =
    -\mathcal{L}_{V}B_{I} +\delta_{\Lambda_{V}}B_{I}
    \\
    & \\
    & =
      -\imath_{V}H_{I}
      +\tfrac{2}{3^{1/2}}C_{IJK}P_{V}{}^{J}F^{K}
      +d\left(\Lambda^{(1)}{}_{V\, I} -\imath_{V}B_{I}
      -\tfrac{1}{3^{1/2}}C_{IJK}P_{V}{}^{J}A^{K}\right)\,,
  \end{aligned}
\end{equation}

\noindent
where we have used the electric momentum map equation
(\ref{eq:electricmomentummapequations1}) with $\mathfrak{h}^{(1)\, I}=0$.
Using the duality relation between the field strengths, we arrive at

\begin{equation}
  -\imath_{V}\left(a_{IJ}\star F^{J}\right)
  +\tfrac{2}{3^{1/2}}C_{IJK}P_{V}{}^{J}F^{K}
  +d\left(\Lambda^{(1)}{}_{V\, I} -\imath_{V}B_{I}
    -\tfrac{1}{3^{1/2}}C_{IJK}P_{V}{}^{J}A^{K}\right)\,,
\end{equation}

\noindent
and, upon use of the magnetic momentum map equation (\ref{eq:magneticmomentummapequation})
we arrive at

\begin{equation}
d\left[\Lambda^{(1)}{}_{V\, I} -\left(\imath_{V}B_{I}-\tilde{P}_{V\, I}\right)
    -\tfrac{1}{3^{1/2}}C_{IJK}P_{V}{}^{J}A^{K}\right]
  +\mathfrak{h}^{(2)}_{V\, I}
  =
  0\,.
\end{equation}

This equation requires the absence of  $\mathfrak{h}^{(2)}_{V\, I}$ in the
magnetic momentum map equations and the identification of the compensating
gauge transformations

\begin{equation}
  \Lambda^{(1)}{}_{V\, I}
  =
  \left(\imath_{V}B_{I}-\tilde{P}_{V\, I}\right)
    +\tfrac{1}{3^{1/2}}C_{IJK}P_{V}{}^{J}A^{K}\,.  
\end{equation}

On the other hand, we can keep the $\mathfrak{h}^{(2)}_{V\, I}$ if we
only demand invariance of the 2-forms $B_{I}$ under the GCT generated
by $V$ up to a global higher-form-type transformation of the form

\begin{equation}
  \delta_{\epsilon}B_{I}
  =
  \epsilon \mathfrak{h}^{(2)}{}_{I}\,.
\end{equation}

However, since the $B_{I}$s do not occur explicitly as fields of the
theory, this does not pose any problems, as different from what
happens with the gauge fields $A^{I}$ and the 1-forms
$\mathfrak{h}^{(1)\, I}$.

The electric momentum maps are defined up to an additive constant. If we
replace $P_{V}{}^{I}$ by

\begin{equation}
  \label{eq:Pprime}
P_{V}{}^{I\, \prime}=P_{V}{}^{I}-\alpha^{I}\,,
\end{equation}

\noindent
where the $\alpha^{I}$ are constant, in the magnetic momentum map
equation (\ref{eq:magneticmomentummapequation}) we find that

\begin{equation}
  \label{eq:tildePprime}  
d\tilde{P}_{V\, I}{}^{\prime}+\mathfrak{h}^{(2)}_{V\, I}{}^{\prime}
=
d\tilde{P}_{V\, I}+\mathfrak{h}^{(2)}_{V\, I}
-\tfrac{2}{3^{1/2}}C_{IJK}\alpha^{J}F^{K}\,.
\end{equation}

$\tilde{P}_{V\, I}{}^{\prime}$ will capture the exact pieces of the
field strengths $F^{I}$ while $\mathfrak{h}^{(2)}_{V\, I}{}^{\prime}$
will capture the closed but non-exact pieces.

Taking the interior product of both sides of the magnetic momentum map
equation (\ref{eq:magneticmomentummapequation}) with $V$ and, under the those
assumptions, we find

\begin{equation}
  d\left(\imath_{V}\tilde{P}_{V\, I}
    -\tfrac{1}{3^{1/2}}C_{IJK}P_{V}{}^{J}P_{V}{}^{K}\right)
\doteq
\imath_{V}\mathfrak{h}^{(2)}_{V\, I}\,,
\end{equation}

\noindent
which implies that

\begin{equation}
  \imath_{V}\mathfrak{h}^{(2)}_{V\, I}
  \doteq
  d\mathbf{M}_{V\, I}\,,
\end{equation}

\noindent
and

\begin{equation}
  d\left(\imath_{V}\tilde{P}_{V\, I}
    -\tfrac{1}{3^{1/2}}C_{IJK}P_{V}{}^{J}P_{V}{}^{K}- \mathbf{M}_{V\, I}\right)
\doteq
0\,.
\end{equation}

We conclude that

\begin{equation}
  \label{eq:iVtildeP}
\imath_{V}\tilde{P}_{V\, I}
-\tfrac{1}{3^{1/2}}C_{IJK}P_{V}{}^{J}P_{V}{}^{K}
-\mathbf{M}_{V\, I}
\doteq
\beta_{I}\,,
\end{equation}

\noindent
for some constants $\beta_{I}$ that we could also have absorbed into the
definition of the functions $\mathbf{M}_{V\, I}$, when the 2-forms
$\mathfrak{h}^{(2)}_{V\, I}$ are not zero.


\end{document}